\documentclass[%
aip,
amsmath,
amssymb,
reprint,
]{revtex4-2}

\usepackage{graphicx}
\usepackage{dcolumn}
\usepackage{bm}

\usepackage[utf8]{inputenc}
\usepackage[T1]{fontenc}
\usepackage{mathptmx}
\usepackage{multirow}
\usepackage{hyperref}
\usepackage{braket}
\usepackage{float}
\usepackage{adjustbox}
\usepackage[version=3]{mhchem} 

\newcommand{\fig}{Fig.}
\newcommand{\Fig}{Fig.}
\newcommand{\figref}[1]{\fig~\ref{#1}}
\newcommand{\Figref}[1]{\Fig~\ref{#1}}

\newcommand{\Tabref}[1]{Table~\ref{#1}}

\renewcommand{\eqref}[1]{Eq.~(\ref{#1})}

\newcommand{\secref}[1]{Section~\ref{#1}}
\newcommand{\Secref}[1]{Section~\ref{#1}}

\newcommand{\corr}[1]{#1} 

\providecommand{\e}[1]{\ensuremath{\times 10^{#1}}}

\newcommand{\thmax}{\theta_{\mathrm{max}}}

\begin{document}

\preprint{}

\title[Predicting Properties of Periodic Systems from Cluster Data: A Case Study of Liquid Water]{Predicting Properties of Periodic Systems from Cluster Data: A Case Study of Liquid Water}

\author{Viktor Zaverkin}
\affiliation{Institute for Theoretical Chemistry, University of Stuttgart, Pfaffenwaldring 55, 70569 Stuttgart, Germany}

\author{David Holzmüller}
\affiliation{Institute for Stochastics and Applications, University of Stuttgart, Pfaffenwaldring 57, 70569 Stuttgart, Germany}

\author{Robin Schuldt}
\affiliation{Institute for Theoretical Chemistry, University of Stuttgart, Pfaffenwaldring 55, 70569 Stuttgart, Germany}

\author{Johannes Kästner}
\email{kaestner@theochem.uni-stuttgart.de}
\affiliation{Institute for Theoretical Chemistry, University of Stuttgart, Pfaffenwaldring 55, 70569 Stuttgart, Germany}

\date{\today}

\begin{abstract}
\corr{The accuracy of the training data limits the accuracy of bulk properties from machine-learned potentials. For example,} hybrid functionals or wave-function-based quantum chemical methods are readily available for cluster data but effectively out-of-scope for periodic structures. We show that local, atom-centred descriptors for machine-learned potentials enable the prediction of bulk properties from cluster model training data, agreeing reasonably well with predictions from bulk training data.  We demonstrate such transferability by studying structural and dynamical properties of bulk liquid water with density functional theory and have found an excellent agreement with experimental as well as theoretical counterparts.

\end{abstract}

\maketitle

\section{\label{sec:intro} Introduction}

The ability to perform simulations that allow for bond formation and bond breaking in aqueous environments is necessary to understand several chemical processes.\cite{Marsalek2017} While first-principles sampling can provide some insight, it is usually limited by its high computational cost. Approximate methods allow for simulations taking into account the large size of conformational and chemical space of interest. Combining molecular dynamics simulations with machine-learned potentials (MLPs), which have been proven to learn any complex non-linear relationship, enables us to extend the simulation time scales and system sizes. These factors are pivotal for obtaining the relevant time correlation functions to converge various dynamical properties in solution, ranging from diffusion coefficients and reorientation times to reaction rates and spectroscopy. 

The advent of machine learning methods in computational chemistry and physics made it feasible to construct efficient interatomic potentials with accuracy on par with reference ab initio methods.\cite{Behler2007, Behler2010, Bartok2010, Rupp2012, Bartok2013, Lilienfeld2015, Shapeev2016, Khorshidi2016, Schuett2017, Faber2018, Kocer2019, Zhang2019, Christensen2020, Zaverkin2020, Schuett2021, Artrith2016, Chmiela2017, Gubaev2018, Lubbers2018, Yao2018, Unke2019, Cooper2020, Zaverkin2021b} Neural networks (NNs), especially, have been proven to approximate any non-linear functional relationship.\cite{Hornik1991}
This fact has promoted their popularity in computational chemistry and materials science.\cite{Morawietz2016, Brickel2019, Kaeser2020, Molpeceres2020, Tovey2020, Korotaev2020, Calegari2020, Molpeceres2021, Zhang2021} NNs were initially applied to represent potential energy surfaces (PESs) of small atomistic systems\cite{Blank1995, Lorenz2004} and were later extended to high-dimensional systems.\cite{Behler2007} Once trained, the computational cost of machine-learned potentials (MLPs) based on NNs is independent of the number of data points used for training as opposed to kernel-based models.\cite{Bartok2010, Rupp2012, Bartok2013, Lilienfeld2015}

\begin{figure*}
\centering
\includegraphics[width=16cm]{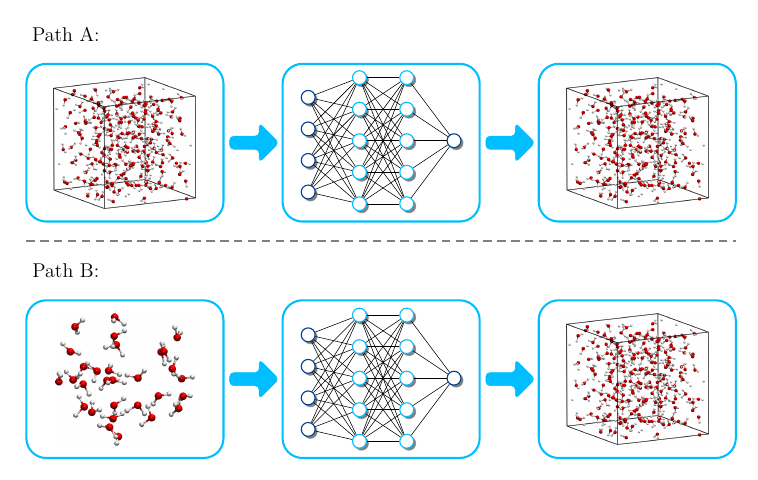}
\caption{Schematic illustration of conventional and investigated physicochemical property prediction path for a periodic system. Path A: An interatomic potential for a periodic system is created using periodic reference data. Path B: An interatomic potential for a periodic system is built using cluster reference data.}
\label{fig:1}
\end{figure*}

Typically, MLPs are trained following path A in \figref{fig:1}, i.e. models supposed to predict periodic systems use periodic reference data for training and vice versa. \corr{Special reference ab initio methods might be required for some applications}, e.g., hybrid functionals such as B3LYP\cite{Becke1993, Stephens1994} or wave-function-based approaches, like coupled-cluster methods. For these, the generation of training data sets for periodic systems can be computationally prohibitive. Taking into account the local, atom-centred architecture of most machine learning approaches employed in computational chemistry and physics, it is, in principle, possible to train the models on cluster data and use them at the inference step for the prediction of periodic systems, see path B in \figref{fig:1}. Such an approach aims to reduce the computational cost for the construction of highly accurate machine-learned interatomic potentials for periodic systems.

Previous work\cite{Ghasemi2015, Faraji2017} employed an architecture that computes total energies and, thus, atomic forces from a charge equilibration process.\cite{Rappe1991, Wilmer2012, Cheng2014} However, here we envisage studying the direct transferability of properties obtained employing the locality assumption from small clusters to bulk materials, rather than using them for an intermediate step to get the global charge density determined by atomic environment-dependent electronegativities.\cite{Ghasemi2015, Faraji2017} \corr{Recently, the transferability of interatomic MLPs has attracted much interest by the community,\cite{Gastegger2016, Monserrat2020, Rowe2020, Westermayr2020, Schran2021, Hajibabaei2021} resulting in an excellent generalisation ability of some MLPs on the out-of-sample configurations. A few cases\cite{Gastegger2016, Schran2021, Hajibabaei2021} demonstrate the transferability of models trained on smaller molecular fragments to larger atomistic systems. In Ref.~\onlinecite{Gastegger2016}, in particular, NN-based MLPs, for which the total energy of a system is built from the environment-dependent atomic energies, have been compared to  fragmentation-based approaches.\cite{Collins2006, Netzloff2007, Collins2012, Gordon2012, Collins2015}}

\corr{In this work,} we investigate the transferability of MLPs trained on local reference data (water clusters) to a periodic liquid water system. \corr{Specifically, we use the open-source package Gaussian moment neural network (GM-NN) for constructing MLPs and applying them in atomistic simulations. The GM-NN source code is available free-of-charge from \href{https://gitlab.com/zaverkin\_v/gmnn}{gitlab.com/zaverkin\_v/gmnn}. Note that no difference between MLPs for water with and without explicit long-range electrostatics has been found for a cutoff radius of $6.35$~\AA{}.~\cite{Morawietz2016} Therefore, all models used in this work are built employing a cutoff of $6.5$~\AA{} but do not contain the explicit long-range electrostatics.}

\corr{To assess the transferability of cluster models, we} thoroughly benchmark \corr{them} comparing the mean absolute errors and respective error distributions to the respective results obtained by employing models trained on periodic data. A more rigorous \corr{evaluation} of the transferability of local models is achieved by running long molecular dynamics \corr{(MD)} simulations in canonical (NVT) statistical ensemble at 300~K. As a baseline for the assessment, we consider structural properties such as oxygen--oxygen, oxygen--hydrogen, and hydrogen--hydrogen radial distribution functions. Additionally, we compute the water self-diffusion coefficient employing local and periodic MLPs. \corr{Besides the NVT simulations, the equilibrium density of liquid water at ambient conditions (at a temperature of 300~K and a pressure of 1~bar) has been estimated by running NPT (constant pressure) simulations.} All results are carefully compared to those obtained from training on bulk data and to experiment.

\corr{While we apply NN-based interatomic potentials to the example of liquid water, many conventional force fields exist to perform accurate simulations of the latter. One of the recent examples is the TL4P water model~\cite{Troester2013} which has been designed to describe the condensed-phase and spectroscopic properties of liquid water correctly.~\cite{Koner2020} However, there are other non-conventional force fields,~\cite{Koner2020} which allow building bulk water models from small clusters.~\cite{Huang2006, Bukowski2007, Babin2012, Babin2013, He2014, Liu2018a, Liu2018b} For a detailed overview of non-conventional force fields, see elsewhere.~\cite{Koner2020} Here, we compare the estimated properties of bulk liquid water by employing cluster models to the fragment-based coupled-cluster approach~\cite{Liu2018a} and the well-established TIP4P water model.\cite{Jorgensen1998} However, the main goal of this work is to investigate the transferability of MLPs trained on cluster data to periodic systems. Therefore, a detailed comparison of their accuracy in terms of predicted energy and forces is out of the scope of this work and the reader is referred to, e.g., Ref.~\onlinecite{Gastegger2016}.}

\section{\label{sec:comp_details} Computational Details}

\subsection{\label{sec:data} Data Sets}

This work demonstrates that atom-centred machine-learned potentials trained on cluster data can be transferred to periodic systems without loss of generalisation capability. For this purpose, we employ two kinds of data sets. The first data set contains structures, energies, and atomic forces of a periodic water box with 64 water molecules (192 atoms) at the revPBE-D3/PAW level of theory.\cite{Bloechl1994, Zhang1998, Grimme2010}
This data set was generated by running an ab initio molecular dynamics (AIMD) simulation at a temperature of 400~K.\cite{Cooper2020} We refer to this data set and the respective interatomic potentials as revPBE-D3\corr{-bulk}. The respective data set is available free-of-charge from Ref.~\onlinecite{Cooper2020_2}.

The second kind of data contains larger water clusters extracted from the N-ASW data set, \corr{created to study nitrogen atoms' adsorption and desorption dynamics} on top of amorphous solid water (ASW).\cite{Molpeceres2020} The N-ASW data set is available free-of-charge from Ref.~\onlinecite{Molpeceres2020_2}. In total, we extracted 1520 structures from the original N-ASW data set: 800 structures with 30 \ce{H2O} sampled at 150 K, 400 structures with 30 \ce{H2O} sampled at 800 K, 200 structures with 60 \ce{H2O} sampled at 300 K, 100 structures with 93 \ce{H2O} sampled at 300 K, 10 structures with 109 \ce{H2O} sampled at 300 K, and 10 structures with 126 \ce{H2O} sampled at 300 K. 

To allow for a comparison with the models trained on the revPBE-D3\corr{-bulk} data set, energies and atomic forces of the respective 1520 configurations were recalculated at the revPBE-D3/def2-TZVP level of theory\cite{Bloechl1994, Zhang1998, Grimme2010, Weigend05, Weigend06} using Turbomole 7.4\cite{Furche2014} within ChemShell.\cite{Metz2014, Sherwood2003} We refer to the respective interatomic potentials and the data set as revPBE-D3\corr{-cluster}. In this work, we are especially interested in using hybrid DFT functionals for the construction of training data which is typically infeasible for periodic structures with a reasonable effort, even for liquid water. Therefore, we compute energies and atomic forces at the B3LYP-D3/def2-TZVP level of theory,\cite{Becke1993, Stephens1994, Grimme2010, Weigend05, Weigend06} popular in the computational chemistry community and refer to the respective data set as B3LYP-D3\corr{-cluster}. In addition to the mentioned functionals, we construct a data set employing the BLYP-D3/def2-TZVP level of theory\cite{Becke1993, Lee88, Grimme2010, Weigend05, Weigend06} and refer to it as BLYP-D3\corr{-cluster}. \corr{The respective data sets are available free-of-charge from Ref.~\onlinecite{Zaverkin2022b}.}

\subsection{\label{sec:mlp} Construction of MLPs}

To study the transferability of atom-centred machine-learned potentials (MLP) trained on cluster data to periodic structures, we employ molecular dynamics (MD) simulations in combination with the Gaussian moment neural network (GM-NN) proposed in Refs.~\onlinecite{Zaverkin2020, Zaverkin2021b}. GM-NN potentials are atom-centered interatomic potentials for which the total energy $E$ of a system $S = \{\mathbf{r}_i, Z_i\}_{i=1}^{N_\mathrm{at}}$, with $\mathbf{r}_i \in \mathbb{R}^3$ being the Cartesian coordinates and $Z_i \in \mathbb{Z}$ being the atomic numbers, is decomposed into its atomic contributions $E_i$
\begin{equation}
    E \left( S, \boldsymbol{\theta} \right) = \sum_{i=1}^{N_\mathrm{at}} E_i\left(\mathbf{G}_i, \boldsymbol{\theta}\right).
\end{equation}
GM-NN models use neural networks (NNs) to map symmetry-preserving local atomic descriptors $\mathbf{G}_i$, Gaussian moments (GMs), to auxiliary atomic energy and include both the geometric and alchemical information about the atomic species of both the central and neighbour atoms. Only a single NN has to be trained for all atomic energy contributions, in contrast to using an individual NN for each species as frequently done in the literature.\cite{Behler2007}

In this work, we use a neural network with two hidden layers consisting of $512$ nodes each and an input dimension of $360$ (number of invariant features). We minimise the combined loss function
\begin{equation}
\label{eq:2}
\begin{split}
\mathcal{L}_{E, \mathbf{F}}\left(\boldsymbol{\theta}\right) = \sum_{k=1}^{N_\mathrm{Train}} & \Bigg[\lambda_E \lVert E_k^\mathrm{ref} - E(S_k, \boldsymbol{\theta})\rVert_2^2 +  \\ & \quad  \frac{\lambda_F}{3N_\mathrm{at}^{(k)}} \sum_{i=1}^{N_\mathrm{at}^{(k)}} \lVert \mathbf{F}_{i,k}^\mathrm{ref} - \mathbf{F}_i\left(S_k, \boldsymbol{\theta}\right)\rVert_2^2\Bigg],
\end{split}
\end{equation}
to optimize the respective parameters of trainable representation, fully-connected neural network part, and the parameters which scale and shift the output of the neural network. For more details on the exact architecture of the network, on the atom-centered representation, and on the definition of energy scale and shift parameters see Ref.~\onlinecite{Zaverkin2021b}.

In \eqref{eq:2}, $N_\mathrm{at}^{(k)}$ is the number of atoms in the respective structure. The reference values for the energy and atomic force are denoted by $E_k^\mathrm{ref}$ and $\mathbf{F}_{i,k}^\mathrm{ref}$, respectively. Since atomic forces containing $3N_\mathrm{at}$ scalars provide much more information about a structure and determine the dynamics of a system, the parameters $\lambda_E$ and $\lambda_F$ were set to $1~\text{au}$ and $12N_\mathrm{at}^{(k)}$~au~{\AA}$^2$, respectively. The loss function in \eqref{eq:2} was minimised employing the Adam optimiser\cite{Adam2015} with $32$ molecules per mini-batch. The layer-wise learning rate was set to $0.03$ for the parameters of the fully connected layers, $0.02$ for the trainable representation, 0.05 and 0.001 for the shift and scale parameters of atomic energies, respectively. All learning rates were decayed linearly to zero. For more information, see Ref.~\onlinecite{Zaverkin2021b}. Besides the hyper-parameters defined above, we have used a cutoff radius of 6.5~\AA{} employed in the definition of the GM representation, see Ref.~\onlinecite{Morawietz2016}.

To run MD simulations with the GM-NN interatomic potentials, we interfaced them with the ASE package (v. 3.21.0).\cite{Hjorth2017} \corr{To assess the accuracy of GM-NN during an MD simulation, its uncertainty has been computed by employing the ensembling technique.\cite{Gastegger2017, Smith2018, ZhangLinfeng2019, Janet2019, Schran2020a, Schran2020b, Imbalzano2021} Note that this method is often referred to as the query-by-committee (QbC) approach\cite{Settles2009} and is used to run active learning, different to the current work. Here,} we trained a committee of 3 models on the same split of the data, i.e. on the same training data set, but using randomly initialised parameters and reported on the respective uncertainty
\begin{equation}
\label{eq:uncertainty}
    \sigma_\mathrm{ens}\left(S\right) = \sqrt{\frac{1}{N_\mathrm{ens}}\sum_{i=1}^{N_\mathrm{ens}} \left(y_i\left(S\right) - \bar{y}\left(S\right)\right)^2},
\end{equation}
where $N_\mathrm{ens}$ is the number of models in the committee, generally we use $N_\mathrm{ens} =3$. $\bar{y}\left(S\right) = 1/N_\mathrm{ens}\sum_{i=1}^{N_\mathrm{ens}} y_i\left(S\right)$ is the mean of the property prediction, the atomic force component in this specific case, over the committee. All models were trained within the Tensorflow framework\cite{Abadi2015} on an NVIDIA Tesla V100-SXM-32GB GPU.

\subsection{Stress Tensor}

\corr{The molecular dynamics (MD) simulations at constant pressure (NPT) in \Secref{sec:density} require virial stress tensors to compute the instantaneous pressure.\cite{Thompson2009} Thus, we have to derive an expression for stress tensors specific to the atom-centred neural networks (NNs) employed in this work. Recall the original expression derived earlier for arbitrary pair potentials\cite{Thompson2009} and used more recently for MLPs\cite{Zhang2018, Shao2020}
\begin{equation}
    \mathbf{W} = -\frac{1}{2} \sum_{i=1}^{N_\mathrm{at}} \sum_{j=1}^{N_\mathrm{at}} \mathbf{r}_{ij} \otimes \mathbf{F}_{ij},
\end{equation}
where $\otimes$ denotes the tensor product, $\mathbf{F}_{ij}$ is the force on atom $i$ caused by the interaction with atom $j$, and $\mathbf{r}_{ij} = \mathbf{r}_i - \mathbf{r}_j$ is the atomic distance vector. Here, $\mathbf{F}_{ij}$ is defined as
\begin{equation}
    \label{eq:force_ij}
    \mathbf{F}_{ij} = -\boldsymbol{\nabla}_{\mathbf{r}_{j}} E_i,
\end{equation}
where $E_i$ is the atomic energy computed by an atom-centered MLP and $\mathbf{r}_j$ are the coordinates of atom $j$.
}

\corr{The atomic energy in the Gaussian moment neural network (GM-NN) framework is defined by the atomic distance vectors, i.e., $E_i = E_i\left(\{\mathbf{r}_{ij}\}_{j \in r_\mathrm{max}}\right)$. Here, $r_\mathrm{max}$ denotes the cutoff radius. Thus, employing the chain rule, the expression in \eqref{eq:force_ij} can be simplified to
\begin{equation}
    \label{eq:gmnn_force_ij}
    \mathbf{F}_{ij} = \boldsymbol{\nabla}_{\mathbf{r}_{ij}} E\left(S, \boldsymbol{\theta}\right),
\end{equation}
where $E\left(S, \boldsymbol{\theta}\right) = \sum_{i=1}^{N_\mathrm{at}} E_i$ is the total energy of an atomic arrangement. The resulting expression for the stress tensor reads
\begin{equation}
    \mathbf{W}(S, \boldsymbol{\theta}) = - \sum_{i=1}^{N_\mathrm{at}} \sum_{j \in r_\mathrm{max}}
        \mathbf{r}_{ij} \otimes \boldsymbol{\nabla}_{\mathbf{r}_{ij}} E(S, \boldsymbol{\theta}).
    \label{eq:7}
\end{equation}
Here, we skip the pre-factor $1/2$ since the second sum in Equation~(\ref{eq:7}) runs only over the local neighbourhood of atom $i$. Therefore, we do not have to account for the double-counting of interactions. Periodic boundary conditions are taken into account when computing $\mathbf{r}_{ij}$. The atoms within the cutoff radius can originate from the
image atoms and the atoms within the cell.}

\section{\label{sec:results} Results and Discussion}

\subsection{\label{sec:mlp_results} Interatomic Neural Network Potentials}

In this section, we discuss the possibility of transferring the interatomic potentials trained on cluster data to periodic systems based on simple error estimates such as absolute error distribution in atomic forces as well as the respective directional error. However, prior to that, the performance of the GM-NN interatomic potentials on each data set in terms of their mean absolute error (MAE) is studied.

\begin{figure*}
\centering
\includegraphics[width=16cm]{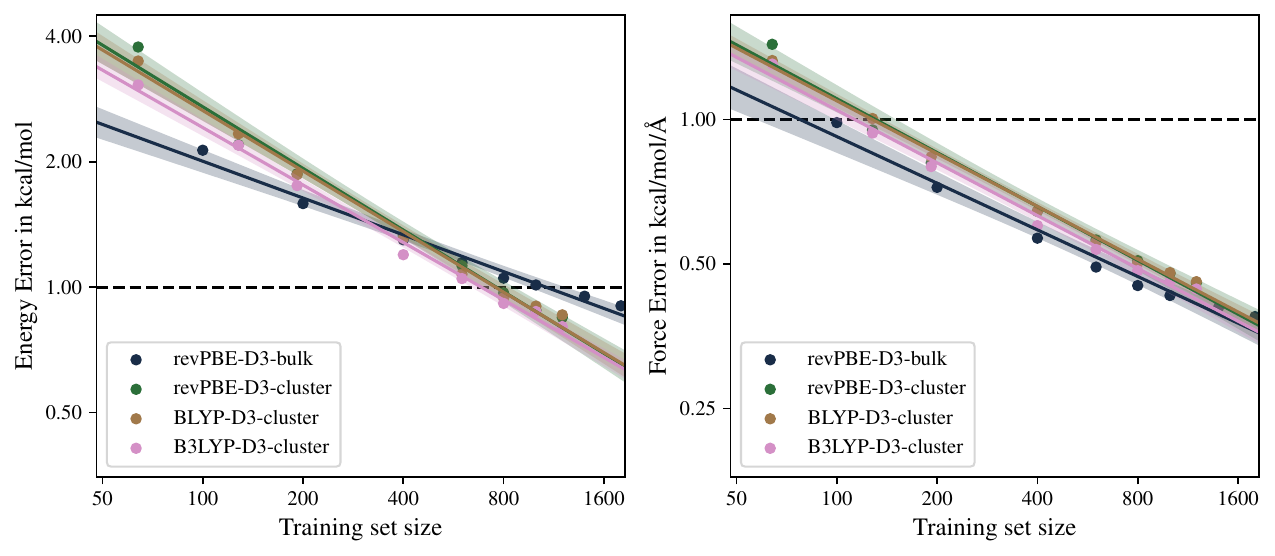}
\caption{Learning curves for periodic and local water data sets. The mean absolute errors (MAEs) of total energies and atomic forces are plotted against the training set size. Linear fits are displayed for clarity, and shaded areas denote the 95 \% confidence intervals for linear regression. The dashed black lines represent the desired accuracies of 1~kcal/mol and 1~kcal/mol/\AA{}, respectively.}
\label{fig:2}
\end{figure*}

\figref{fig:2} shows the learning curves for models trained on periodic revPBE-D3\corr{-bulk} and local revPBE-D3\corr{-cluster}, BLYP-D3\corr{-cluster}, B3LYP-D3\corr{-cluster} data sets. From the figure, we see that the GM-NN interatomic potentials trained on revPBE-D3\corr{-bulk} reach the desired accuracy of 1 kcal/mol for total energies already for a training set size of about 1000 reference structures. The desired accuracy of 1 kcal/mol/\AA{} for atomic forces is reached already after training on 100 reference structures. A similar trend can be observed for GM-NN potentials trained on local data sets. However, an MAE of 1~kcal/mol in total energies is reached slightly earlier when training on about 800 configurations.

To test the transferability of the GM-NN interatomic potentials trained on local data to periodic systems, we predict total energies and atomic forces using the local GM-NN models for the periodic revPBE-D3\corr{-bulk} test data and compare the acquired results with the predictions of periodic GM-NN models. Note that we perform a direct comparison only for the local models trained on the revPBE-D3\corr{-cluster} data set since for the other local data sets, similar results are expected. 

\begin{figure*}
\centering
\includegraphics[width=16cm]{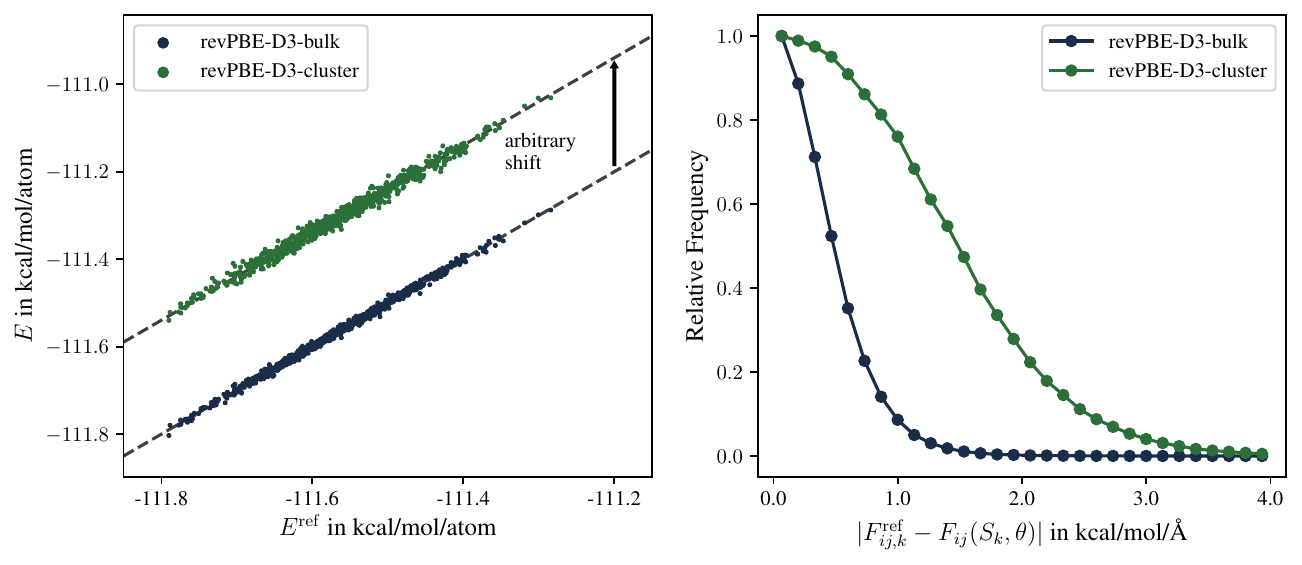}
\caption{(Left) Correlation of the predicted potential energies with the corresponding reference values for all structures in the revPBE-D3\corr{-bulk} test data. An arbitrary energy shift is introduced to the energies predicted by the GM-NN model trained on local revPBE-D3\corr{-cluster} data to allow for a comparison with the models trained on the periodic revPBE-D3\corr{-bulk} data. (Right) Distribution of the absolute force error for GM-NN potentials trained on local and periodic data. Here, $F_{ij,k}$ is the force component $j \in \{1, 2, 3\}$ acting on the atom $i \in \{1, \dots, N_\mathrm{at}^{(k)}\}$ in the structure $k \in \{1, \dots, N_\mathrm{test}\}$. Periodic models were trained using 1800 reference structures; local models were trained on 1400 reference structures.}
\label{fig:3}
\end{figure*}

\Figref{fig:3} (left) shows the correlation of the GM-NN potential energies with the corresponding ab initio reference energies obtained for the structures in the revPBE-D3\corr{-bulk} test data split. As seen in the figure, the GM-NN models trained on data containing only water clusters can reproduce the total energy of a periodic bulk water system with 64 water molecules (192 atoms) up to an arbitrary shift. The latter is introduced mainly by the differences in ab initio methods employed in \secref{sec:data}. Note that we shifted the revPBE-D3\corr{-cluster}-based GM-NN potential energies in \figref{fig:3} (left) to allow for a better comparison with revPBE-D3\corr{-bulk}-based GM-NN predictions.

\Figref{fig:3} (right) shows the distribution of the absolute error in the components of the predicted atomic forces for the structures within the revPBE-D3\corr{-bulk} test data set, employing MLPs trained on the local and periodic data. From the figure, one can see that the distribution is broader for the MLPs trained on local data compared to the models trained on periodic data. However, that may originate from the differences in levels of theory and other settings used to generate the training data. Additionally, MLPs trained on periodic data can explicitly model the interactions with images from neighbouring cells, while local models are not trained to do so. Therefore, we study this phenomenon more rigorously, e.g., by evaluating the error in directions of atomic forces and by evaluating structural and dynamical properties from a molecular dynamics (MD) simulation.

\begin{figure*}
\centering
\includegraphics[width=16cm]{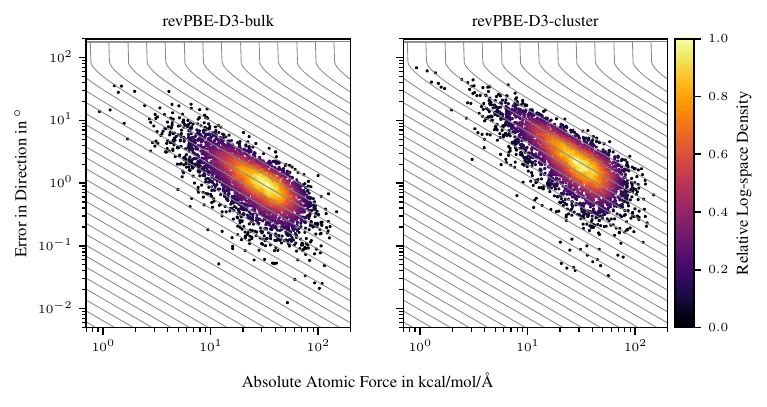}
\caption{Samples of the angular error of the atomic force in degrees versus the absolute atomic force for GM-NN models trained on local revPBE-D3\corr{-cluster} and periodic revPBE-D3\corr{-bulk} data. The colours indicate the relative density in log-space as estimated by a kernel density estimator. The gray lines correspond to $F \mapsto \thmax(E, F)$ (in degrees) for different values of $E$, see \eqref{eq:max_angle}.}
\label{fig:4}
\end{figure*}

\Figref{fig:4} shows a corresponding analysis of the error in the direction of the predicted atomic forces. The plots show, for different atoms in the revPBE-D3\corr{-bulk} test set, the absolute reference force and the angle difference between reference and predicted forces. Brighter colours indicate a higher density of points in the log-log plot. It can be seen that the maximal angular error decreases with increasing absolute values of the force vectors, for both MLPs trained on local and those trained on periodic data.
This can be explained by the loss function trying to minimize the absolute force error. For a given reference absolute force $F = \|\mathbf{F}^{\mathrm{ref}}_{i,k}\|_2$ and given absolute force error $E = \|\mathbf{F}^{\mathrm{ref}}_{i,k} - \mathbf{F}_i(S_k, \mathbf{\theta})\|_2$, the maximum possible angle is
\begin{equation}
    \thmax(E, F) = \begin{cases}
    \arcsin(E/F) &,\text{ if $E \leq F$} \\
    \pi &,\text{ if $E > F$.}
    \end{cases} \label{eq:max_angle}
\end{equation} 
The grey lines in the figure correspond to $F \mapsto \thmax(E, F)$ for different values of $E$. It can be seen that the maximum angles are well-aligned with these curves.

As seen from the figure, the direction of the force vector scatters slightly stronger for the MLP trained on local data compared to the one trained on periodic data. By \eqref{eq:max_angle}, this is in line with the larger force errors shown in \Figref{fig:3}. Since the directional errors are typically still small for the MLP trained on water clusters, it can be expected that such MLPs apply to a broad range of atomistic simulations of periodic liquid water.

\subsection{\label{sec:properties} Water Properties}

The MAE of total energies and atomic forces and the respective error distributions discussed earlier are abstract quality measures of machine-learned interatomic potentials. In practice, the robustness and reliability of the potential in a real-time application are more important. 
Therefore, in this section, we assess the robustness and smoothness of the GM-NN interatomic potentials based on local data sets by applying them to a periodic liquid water system containing 216 water molecules (648 atoms) to run a molecular dynamics (MD) simulation. This is because for an MD simulation, a smooth energy surface is required to obtain an adequate estimate for structural properties such as radial distribution functions and dynamical properties like diffusion coefficients.

In this section, in order to analyse properties of liquid water, we run molecular dynamics (MD) simulations in the canonical (NVT) statistical ensemble carried out within the ASE simulation package\cite{Hjorth2017} using a Langevin thermostat at the temperature of 300~K but different atomic densities depending on the level of theory employed. 
For revPBE-D3\corr{-bulk} and revPBE-D3\corr{-cluster}, we have used the atomic density of liquid water at ambient conditions of $0.998$~g~cm$^{-3}$. For BLYP-D3\corr{-cluster} and B3LYP-D3\corr{-cluster}, we employed slightly higher values of 1.04 and 1.06~g~cm$^{-3}$, obtained by performing several experiments in which the atomic density was varied between $0.998$~g~cm$^{-3}$ and 1.06~g~cm$^{-3}$.
Note that it is known that BLYP-D3-based MD simulations produce slightly denser liquid water,\cite{Gillan2016} and the present study shows that the same can be expected for B3LYP-D3-based simulations.
All MD runs were performed over 2.0~ns using a time step of 0.5~fs. The atomic velocities were initialised with a Maxwell--Boltzmann distribution for a temperature of 300~K.

\begin{figure}
    \centering
    \includegraphics[width=8cm]{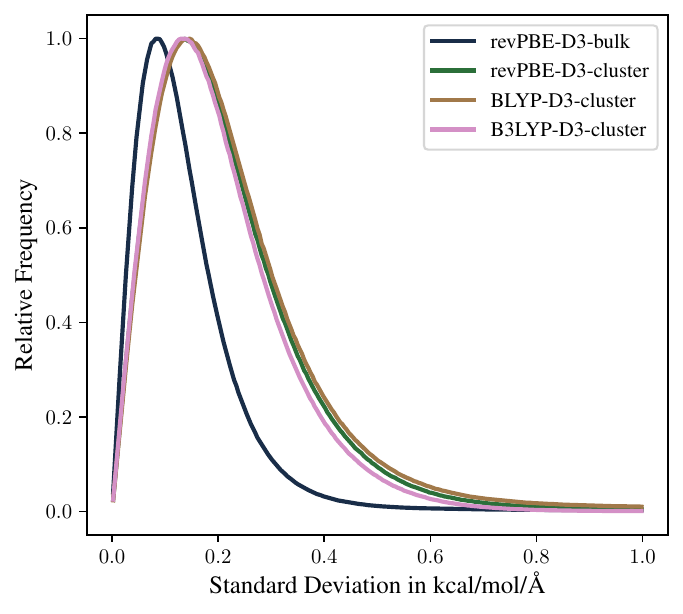}
    \caption{Uncertainty distribution for atomic force components obtained during molecular dynamics simulation on bulk liquid water obtained via the QbC approach\cite{Settles2009} for models trained on periodic revPBE-D3\corr{-bulk} and local revPBE-D3\corr{-cluster}, BLYP-D3\corr{-cluster}, and B3LYP-D3\corr{-cluster} reference data sets.}
    \label{fig:5}
\end{figure}


Forces for molecular dynamics are generated by an ensemble of three machine-learned interatomic potentials, see \eqref{eq:uncertainty}. The ensembling technique\corr{\cite{Gastegger2017, Smith2018, ZhangLinfeng2019, Janet2019, Schran2020a, Schran2020b, Imbalzano2021}} provides us with an error estimate of the potential during simulation. We have found the machine-learned potential to be very accurate with uncertainty between models with a mean ranging from $0.17$~kcal/mol/\AA{} to $0.28$~kcal/mol/\AA{}, see \figref{fig:5}. Due to the cutoff radius, a model cannot directly notice whether periodic boundary conditions are used or not. Hence, the larger uncertainty of local models indicates that the bulk data is locally different from cluster data used during training, requiring the local models to extrapolate. These differences might arise because cluster data contain fewer bulk-like atoms compared to periodic data and also by other aspects of training data generation.
Interestingly, we have observed only a minor increase in uncertainty of a factor of a maximum of 1.6, leading to a still well-behaving model with a mean uncertainty of $0.28$~kcal/mol/\AA{}. These observations suggest that the models trained on local data do not reach a strongly extrapolative region, making them suitable for the following study of the liquid water system.

\corr{Note that an ensemble of three models can be deemed too small, and larger uncertainties could be expected. The respective values can be re-scaled to get more accurate estimates; see Ref.~\onlinecite{Imbalzano2021}. However, one should consider that the uncertainty generally does not have to correlate with the actual error, i.e., a large uncertainty does not necessarily imply a large error. For example, in Ref.~\onlinecite{Zaverkin2021a}, an upper bound for the correlation of uncertainty of atomistic NNs with the actual error has been computed. Here, the estimated uncertainties should only shed light on the relative stability of MLPs during MD simulations. A more rigorous assessment of the quality of trained MLPs is achieved by analyzing the estimated properties, i.e., the radial distribution functions and diffusion coefficients.}

\subsubsection{\label{sec:rdf} Radial Distribution Function}

In the following, our discussion on liquid water will be concerned with the structure of bulk liquid water and the dynamics of the water molecules. We shall pay particular attention to the three radial distribution functions (RDFs) $g_{\alpha \beta}\left(r\right)$ and the self-diffusion coefficient $D$, all of which are available from experiments. All these properties are obtained from the aforementioned 2.0~ns long MD simulations in the canonical (NVT) ensemble after the 20.0~ps equilibration period. We split the whole trajectory into ten equally-sized parts to get statistics on the estimated RDF curves.

\begin{figure*}
\centering
\includegraphics[width=15cm]{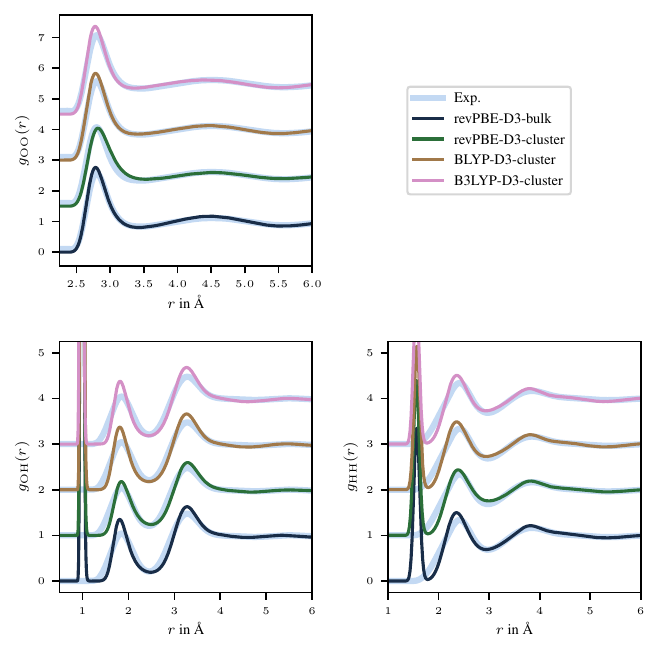}
\caption{Oxygen--oxygen ($g_{\ce{OO}}$), oxygen--hydrogen ($g_{\ce{OH}}$), hydrogen--hydrogen ($g_{\ce{HH}}$) radial distribution functions (RDF) of liquid water at ambient conditions from experiment and from MD simulations based on GM-NN models trained on periodic revPBE-D3\corr{-bulk}, local revPBE-D3\corr{-cluster}, BLYP-D3\corr{-cluster}, and B3LYP-D3\corr{-cluster} data. Experimental results are taken from Refs.~\onlinecite{Skinner2013, Soper2013} for $g_{\ce{OO}}$, $g_{\ce{OH}}$, and $g_{\ce{HH}}$, respectively.}
\label{fig:6}
\end{figure*}

As an initial orientation, we refer to \figref{fig:6} showing the oxygen--oxygen ($g_{\ce{OO}}$), oxygen--hydrogen ($g_{\ce{OH}}$), and hydrogen--hydrogen ($g_{\ce{HH}}$) RDFs for liquid water at ambient conditions obtained from experiment\cite{Skinner2013, Soper2013} and from MD simulations employing GM-NN interatomic potentials. The experimental data for the oxygen--oxygen ($g_{\ce{OO}}$) RDF originates from high-energy X-ray diffraction measurements performed at temperatures close to 297~K.\cite{Skinner2013}

As shown in \figref{fig:6}, the $g_{\ce{OO}}$ curve simulated employing the revPBE-D3\corr{-bulk}-based potential is in good agreement with the experimental results for both the positions and intensities of the first two peaks, see \Tabref{tab:1}. The respective values obtained from MD simulation are very close to the experimental ones and deviate from them by at most 0.03~\AA{} in positions and 0.20 in intensities. One should note that employing the local variant revPBE-D3\corr{-cluster} leads to an even better agreement of $g_{\ce{OO}}$ with its experimental counterpart, especially considering the higher values of the trough between the first two peaks. The results are also consistent with the ab initio study in Ref.~\onlinecite{Marsalek2017}.

The BLYP-D3\corr{-cluster} and B3LYP-D3\corr{-cluster} based GM-NN potentials produce fairly good $g_{\ce{OO}}$ curves with an excellent minimum between the first two peaks but an overestimated intensity of the first peak. Positions of all peaks are in good agreement with the experiment. Note that such behaviour is known for the BLYP-D3 level of theory,\cite{Gillan2016} and interestingly is also observed for the hybrid B3LYP-D3 functional.

\begin{table*}
\caption{\label{tab:1}Positions and intensities of the 1st maximum $(r_1, g_1)$, 1st minimum $(r_2, g_2)$, and 2nd maximum $(r_3, g_3)$ in $g_{\ce{OO}}$ with the respective standard deviations given in parentheses.
Experimental data is from high-energy x-ray diffraction measurements.\cite{Skinner2013} Positions are given in \AA{}.}
\begin{ruledtabular}
\begin{tabular}{lccccc}
    & Exp. X-ray & revPBE-D3\corr{-bulk} & revPBE-D3\corr{-cluster} & BLYP-D3\corr{-cluster} & B3LYP-D3\corr{-cluster} \\
\hline
$r_1$           & 2.80(1)   & 2.79(1) & 2.82(1) & 2.79(1) & 2.78(1) \\
$g_1$           & 2.57(5)   & 2.77(2) & 2.55(2) & 2.83(2) & 2.87(2) \\
$r_2$           & 3.45(4)   & 3.43(2) & 3.54(3) & 3.41(3) & 3.37(2) \\
$g_2$           & 0.84(2)   & 0.80(2) & 0.88(1) & 0.85(2) & 0.85(1) \\
$r_3$           & 4.5(1)    & 4.47(3) & 4.51(3) & 4.47(3) & 4.42(3) \\
$g_3$           & 1.12(2)   & 1.18(1) & 1.10(1) & 1.13(1) & 1.12(1)
\end{tabular}
\end{ruledtabular}
\end{table*}

All oxygen--oxygen RDFs, experimental and the ones sampled from MD simulations, show a clear considerable disorder, since the $g_{\ce{OO}}$ curve is quite close to unity over the range $3.0 < r_{\ce{OO}} < 4.0$~\AA{}. The diffraction measurements agree that the intensity of $g_{\ce{OO}}$ at its first minimum is $0.84 \pm 0.02$. All the GM-NN-based simulations resulted in only slightly over- or under-structured liquids with the height of the first peak and the first minimum in $g_{\ce{OO}}$ being 2.55--2.87 and 0.80--0.88, respectively. Note that typical values for the revPBE functional range from 2.34 to 2.94 and from 0.76 to 1.10 for $g_1$ and $g_2$, respectively,\cite{Gillan2016} depending on the simulation algorithm and dispersion correction. For the BLYP-based simulations, the respective values cover an even broader range, 2.40--3.17 and 0.52--1.0.\cite{Gillan2016} It shows that the approximation introduced by the GM-NN model does not significantly influence the resulting structural properties sampled during a long MD simulation.

For oxygen--hydrogen ($g_{\ce{OH}}$) and hydrogen--hydrogen ($g_{\ce{HH}}$) curves, the intensities of \corr{all} peaks shown in \figref{fig:6} are slightly overestimated in comparison with the experimental results. It can be attributed to the lack of quantum nuclear effects (QNE) in solutions and is out of the scope of this paper. \corr{We should mention that, here, no direct comparison for the first peaks of the $g_{\ce{OH}}$ and $g_{\ce{HH}}$ curves could be made} since those are not given by the experiment used as a reference in this work.\cite{Soper2013} \corr{However, it is known that they are, typically, very overstructured in classical MD simulations, and the inclusion of QNE is necessary to get correct estimates.\cite{Marsalek2017}} Note that the $g_{\ce{OH}}$ curves for revPBE-D3\corr{-bulk} and revPBE-D3\corr{-cluster} agree with the results obtained for the oxygen--hydrogen RDF in Ref.~\onlinecite{Pestana2017}.

\subsubsection{\label{sec:diffusion} Diffusion Coefficient}

Another physical quantity we are interested in is the self-diffusion coefficient of the water molecules. It can be calculated from a linear fit of the mean square displacement (MSD) of the molecular centre of mass and performing a fit using
\begin{equation}
    \langle \Delta x^2\rangle = 2Dd\Delta t,
\end{equation}
where $D$ is the diffusion coefficient, $d=3$ the dimensionality of the system, and $\Delta t$ the time interval for which the respective MSD was measured. In this work, we used time intervals in the range of 1--10~ps. Because of the finite-size effect on diffusion, an extrapolation to infinite system size is required to allow for comparison to the experiment.\cite{Yeh2004} To get the extrapolated diffusion coefficient $D\left(\infty\right)$ we add a correction to the the finite-system diffusion coefficient $D\left(L\right)$ and the resulting expression reads\cite{Yeh2004}
\begin{equation}
    D\left(\infty\right) = D\left(L\right) + \frac{\xi}{6 \pi \beta \eta L}.
\end{equation}
Here, $L$ is the length of the simulation box, $\beta=1/k_\mathrm{B}T$, $\xi$ is given by the geometry of the simulation box ($\xi = 2.837297$ for a cubic box) and $\eta=0.8925\e{-3}$~Pa~s is the experimental shear viscosity.

\begin{figure}
\centering
\includegraphics[width=8cm]{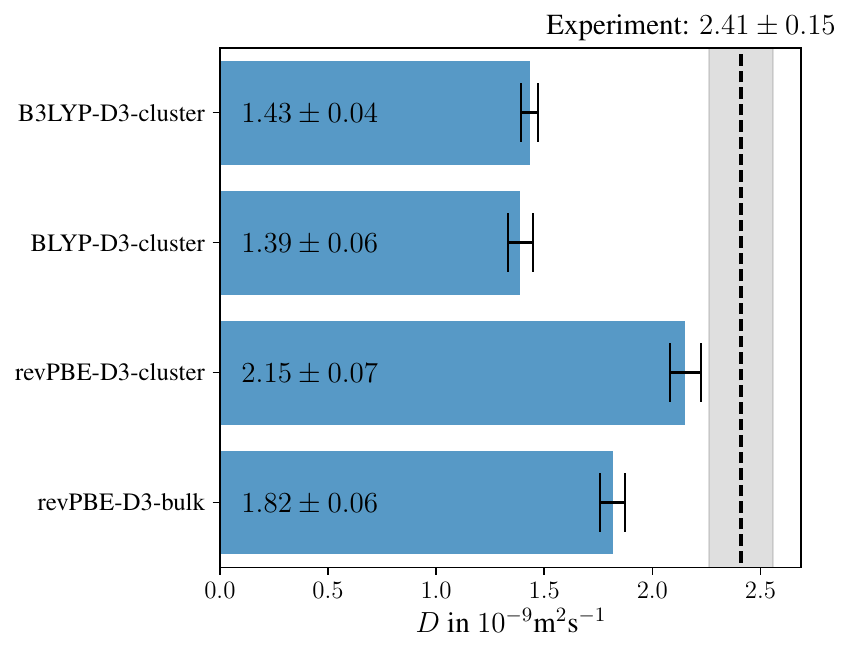}
\caption{Water molecule self-diffusion coefficients from molecular dynamics simulations at 300~K based on revPBE-D3\corr{-bulk}, revPBE-D3\corr{-cluster}, BLYP-D3\corr{-cluster}, and B3LYP-D3\corr{-cluster} GM-NN potentials. Error bars are obtained as the standard deviation between 10 estimates. The black line corresponds to the experimental value while the shaded area denotes its error bar.\cite{Holz2000}}
\label{fig:7}
\end{figure}

\Figref{fig:7} shows the system size corrected diffusion coefficients obtained from the MD simulations based on GM-NN interatomic potentials along with the respective experimental value.\cite{Holz2000} Similar to \secref{sec:rdf} to get some uncertainty estimate for the respective values of diffusion coefficients, we split the whole 2.0~ns long trajectory into ten parts. Error bars for the ML-based diffusion coefficients are obtained as the standard deviation between 10 estimates.

The system size corrected diffusion coefficients of $\left(1.82 \pm 0.06 \right)\e{-9}$~m$^2$~s$^{-1}$, $\left(2.15 \pm 0.07 \right)\e{-9}$~m$^2$~s$^{-1}$, $\left(1.39 \pm 0.06 \right)\e{-9}$~m$^2$~s$^{-1}$, and $\left(1.43 \pm 0.04 \right)\e{-9}$~m$^2$~s$^{-1}$ obtained employing revPBE-D3\corr{-bulk}, revPBE-D3\corr{-cluster}, BLYP-D3\corr{-cluster}, and B3LYP-D3\corr{-cluster}-based interatomic potentials, respectively. All these values are in fair agreement with experimental value of $\left(2.41 \pm 0.15\right)\e{-9}$~m$^2$~s$^{-1}$.\cite{Holz2000} Note that the empirical force fields usually produce a rather larger diffusion coefficient of $>2.6\e{-9}$~m$^2$~s$^{-1}$,\cite{Price2004} while the DFT-based AIMD simulations usually give a relatively smaller value of $<2.0\e{-9}$~m$^2$~s$^{-1}$.\cite{DiStasio2014, Pestana2017} More specifically, the ab-initio MD simulations result in self-diffusion coefficients of merely 0.71\e{-9} to 1.81\e{-9}~m$^2$~s$^{-1}$ and 1.6\e{-9} to 3.4\e{-9}~m$^2$~s$^{-1}$ for BLYP and revPBE, respectively, depending on the simulation algorithm and the dispersion correction,\cite{Gillan2016} and are close to the values predicted using our potentials. 

Using the GM-NN potential trained on local revPBE-D3\corr{-cluster} data produces diffusion coefficients close to the one obtained in Ref.~\onlinecite{Marsalek2017} of $\left(2.22 \pm 0.05\right)\e{-9}$~m$^2$~s$^{-1}$, where AIMD simulations were performed with the revPBE-D3 functional. Moreover, using the local models, we are able to produce results close to the one presented in Ref.~\onlinecite{Liu2018a} ($2.0\e{-9}$~m$^2$~s$^{-1}$) where the fragment-based coupled-cluster approach was employed.

As a final remark, we would like to note that besides the usage of machine-learned interatomic potentials, other approximations made in this study may influence the observed quantities, such as the employed canonical (NVT) statistical ensemble for the system and Langevin thermostat for adjusting the system's temperature. The agreement of the computational results obtained in this work with their experimental counterparts is mainly achieved only due to the accurate atom-centred machine-learned potentials and their ability to transfer from cluster to periodic systems.

\subsection{\label{sec:density} Estimating equilibrium density of bulk liquid water}

\corr{The ability of MLPs trained on cluster data to reproduce the correct equilibrium density of bulk liquid water at ambient conditions is a very useful criterion to assess the generalization ability of the former. For this purpose, we run molecular dynamics (MD) simulations in the NPT (constant pressure) statistical ensemble. Specifically, the isobaric-isothermal form of the Nos{\'e}--Hoover dynamics has been employed\cite{Melchionna1993, Melchionna2000} as implemented in ASE.\cite{Hjorth2017} All production simulations have been performed at ambient conditions, i.e., $T = 300$~K and $p = 1$~bar, over 1~ns after 200~ps long equilibration. A time step of 0.5~fs has been used to integrate the equations of motion, while the characteristic time scales of the thermostat and barostat were set to 1~ps each. The simulation box was allowed to change independently along the three Cartesian axes, $x$, $y$, and $z$. However, the angle between axes has been fixed to 90 degrees. Finally, the density of the bulk liquid water has been computed from the time-averaged (1~ns) volume of the simulation box.}

\corr{The MD simulations for the MLP trained on the periodic revPBE-D3\corr{-bulk} data set resulted in an underestimated density, $0.86$~g~cm$^{-3}$. By contrast, the MLP trained on local revPBE-D3\corr{-cluster} data produced bulk liquid water with a much more realistic  density of $1.02$~g~cm$^{-3}$. The well-established TIP4P water model provides a density of $1.00$~g~cm$^{-3}$ for comparison.\cite{Jorgensen1998} The difference between the estimated densities employing cluster and the periodic models might arise from various aspects of training data generation. Moreover, on par with the observations from \Secref{sec:properties}, BLYP-D3-cluster and B3LYP-D3-cluster-based NN potentials produce bulk liquid water with an overestimated density of $1.12$ and $1.10$~g~cm$^{-3}$, respectively.} 

\corr{We have observed in \Secref{sec:properties}, in agreement with previous  studies,\cite{Gillan2016,Marsalek2017} that the revPBE-D3 level is well suited to describe bulk liquid water. However, the primary purpose of this section is not to assess the reference DFT but rather to study the limitations of the cluster models in terms of their generalisation ability. From the presented results, the excellent transferability of cluster models to periodic bulk systems can be deduced. However, further studies on systems different from bulk liquid water are necessary to get a deeper insight into this intriguing feature of MLPs.}

\corr{Under reasonably general conditions, the only difference between the cluster and bulk models are the local environments of surface atoms and the atoms at the periodic boundaries. The former lack bonding partners and, thus, experience increased strain energy. This effect leads to atomic arrangements at the surface being different from bulk ones. Note that they are physically reasonable and do not bias the model to a nonphysical behaviour. When running simulations of periodic systems with the cluster models, these atomic arrangements are not probable and, thus, are not visited. However, they might contribute to a better performance of cluster models when computing stress tensors. It could explain the better performance of the revPBE-D3-cluster-based model than the revPBE-D3-bulk-based model.}

\corr{Another aspect that could influence the performance of MLPs trained on cluster data are long-range electrostatic interactions, which are absent in our MLPs. However, no important contributions of the long-range interactions could be found, which is in agreement with previous work,\cite{Morawietz2016} where no significant difference in the accuracy between MLPs with and without explicit long-range electrostatics for a cutoff of 6.35~\AA{}. We have employed a cutoff of 6.5~\AA{}.}

\section{\label{sec:conclusions} Conclusions}

We have studied the capability of atom-centred machine-learned potentials (MLPs) to transfer the information gained from local training data to periodic systems. \corr{Such a transfer is essential for the construction of training data sets for periodic systems from ab initio methods not feasible otherwise, e.g., hybrid functionals such as B3LYP. Simular transferability is used for non-conventional force fields other than MLPs, e.g., the recent fragment-based coupled-cluster approach.\cite{Liu2018a} For more examples of non-conventional force fields, see Ref.~\onlinecite{Koner2020} and references therein. Here, the transfer of structure-property predictions of atom-centred MLPs} has been studied on an example of liquid water, while more elaborate applications can be feasible, such as molten alkali chloride salts.\cite{Tovey2020} \corr{Note that no explicit electrostatics for our water models has been employed since it is irrelevant for large enough cutoff radii.\cite{Morawietz2016}}

Applying the machine-learned potentials trained on local data, we could obtain oxygen--oxygen, oxygen--hydrogen, and hydrogen--hydrogen radial distribution functions (RDF). We have found a good agreement with the experiment for the oxygen--oxygen RDF, while for the others, somewhat overestimated intensities were observed due to the lack of quantum nuclear effects (QNE). Moreover, we obtained values reasonably close to the experimental data and on par with the recent ab initio molecular dynamics studies by studying the self-diffusion coefficients. \corr{Finally, by running molecular dynamics simulations at constant pressure, the equilibrium densities for our cluster models could be estimated. For the revPBE-D3-cluster-based model, we have found an excellent agreement with the experiment, while its periodic counterpart provided somewhat underestimated density.}

In summary, we want to \corr{state} the central finding of this work. We have shown that it is feasible to transfer machine-learned interatomic potentials trained on local data to periodic systems. It reduces the computational cost required for the creation of training data using ab-initio methods more suited for cluster models, e.g. hybrid functionals such as B3LYP. Moreover, the possibility to transfer a local machine-learned potential to periodic systems opens the door to predictions of bulk physical quantities using traditional wave-function-based approaches, like coupled-cluster \corr{approach}. However, further development to reach such goals is necessary, which will be a subject of our future investigations.

\begin{acknowledgments}

We thank the Deutsche Forschungsgemeinschaft (DFG, German Research Foundation) for supporting this work by funding EXC 2075 - 390740016 under Germany's Excellence Strategy. We acknowledge the support by the Stuttgart Center for Simulation Science (SimTech). The authors acknowledge support by the state of Baden-Württemberg through the bwHPC consortium for providing computer and GPU time. The authors thank the International Max Planck Research School for Intelligent Systems (IMPRS-IS) for supporting D. H., while V. Z. acknowledges the financial support received in the form of a PhD scholarship from the Studienstiftung  des  Deutschen  Volkes (German National Academic Foundation).

\end{acknowledgments}

\bibliography{aipsamp}

\end{document}